\documentclass[aps,prl,reprint,superscriptaddress,showpacs]{revtex4-1}

\usepackage{graphicx}
\usepackage{amsmath}
\usepackage{mathtools}
\usepackage{float}
\usepackage{color,soul}
\usepackage[pdftex,colorlinks=true,bookmarks=false,citecolor=blue,urlcolor=blue]{hyperref}

\begin{document}

\title{Thermodynamic Effects of Single-Qubit Operations in Silicon-Based Quantum Computing}

\author{Pavel Lougovski}
\email{lougovskip@ornl.gov}
\affiliation{Quantum Information Science Group, Oak Ridge National Laboratory, Oak Ridge, Tennessee 37831, USA}
\author{Nicholas A. Peters}
\affiliation{Quantum Information Science Group, Oak Ridge National Laboratory, Oak Ridge, Tennessee 37831, USA}
\affiliation{The Bredesen Center for Interdisciplinary Research and Graduate Education, The University of Tennessee, Knoxville,  Tennessee 37996, USA }

\date{\today}

\begin{abstract}
Silicon-based quantum logic is a promising technology to implement universal quantum computing. It is widely believed that a millikelvin cryogenic environment will be necessary to accommodate silicon-based qubits. This prompts a question of the ultimate scalability of the technology due to finite cooling capacity of refrigeration systems. In this work, we answer this question by studying energy dissipation due to interactions between nuclear spin impurities and qubit control pulses. We demonstrate that this interaction constrains the sustainable number of single-qubit operations per second for a given cooling capacity. Our results indicate that a state-of-the-art dilution refrigerator can, in principle, accommodate operations on millions of qubits before thermal energy dissipation becomes a problem.\footnote{This manuscript has been authored by UT-Battelle, LLC under Contract No. DE-AC05-00OR22725 with the U.S. Department of Energy.  The United States Government retains and the publisher, by accepting the article for publication, acknowledges that the United States Government retains a non-exclusive, paid-up, irrevocable, world-wide license to publish or reproduce the published form of this manuscript, or allow others to do so, for United States Government purposes.  The Department of Energy will provide public access to these results of federally sponsored research in accordance with the DOE Public Access Plan(
\href{http://energy.gov/downloads/doe-public-access-plan}{http://energy.gov/downloads/doe-public-access-plan}). }
\end{abstract}


\maketitle

%\section{Introduction}

Environmentally-induced decoherence is typically considered as the main obstacle on the path towards a scalable quantum computer~\cite{Ladd,RevModPhys.88.041001}. Known remedies include decreasing qubit coupling to the environment~\cite{lidar14,RevModPhys.88.041001} and using quantum error correcting (QEC) codes~\cite{Gaitan:2007:QEC:1554882,lidar2013quantum}. Then the scalability challenge lies in striking the right balance between the size of QEC and the magnitude of the decoherence effect. What often goes overlooked in this analysis is the effect of qubit controls on the environment itself. While the direct effects of the control pulse coupling to the environment can in principle be minimized by employing shielding, the indirect, or qubit-mediated, coupling of the controls to the environment is likely to be a fundamental challenge. In turn, the latter may result in a net energy increase of the environment which inevitably has to be removed from the system. The magnitude of this effect per qubit, however small, is not negligible. Considering that future quantum computers could utilize tens of millions of physical qubits, even small thermal effects in aggregate could constrain the ultimate scalability. 

In this paper, we study this exact problem for a phosphorus donor nuclear spin qubit in silicon~\cite{Kane}.  We investigate the thermodynamic implications of qubit control pulses coupling with silicon substrate impurities. Our choice of the physical qubit is largely motivated by a multitude of attractive features offered by silicon: long qubit coherence times; a potential for scalable manufacturing via mature CMOS techniques; and tremendous recent experimental advances in controlling a single donor  atom~\cite{Pla2,Pla1,Muhonen,Pla3} or, alternatively, small clusters of atoms~\cite{Buch2013}. The decoherence mechanism is mainly due to the interaction with unavoidable inclusions of $^{29}$Si isotope (nuclear spin $\frac{1}{2}$) that may range in concentration from 4.7$\%$ in natural occurring silicon to below 100 PPM in highly purified samples~\cite{Becker}. The $^{31}{\rm P}$-donor nuclear and electron spin coherence times are well characterized both theoretically ~\cite{YaoLiuSham, Saikin, Cywinski} and experimentally~\cite{Pla2, Tyryshkin, Muhonen}. For example, at 100 mK, in isotopically purified $^{28}$Si, donor nuclear and electron spin  coherence times of up to 30 s~\cite{Muhonen} and 1 s~\cite{Tyryshkin}, respectively, have been demonstrated, potentially supporting thousands or even tens of thousands of quantum gates. Further, the use of topological QEC codes promise to bring fault tolerance within reach~\cite{Hill}. 

One's ability to coherently manipulate the nuclear spin of a $^{31}{\rm P}$ donor is important not only for single qubit operations but also for two-qubit gates between neighboring donor electron spin qubits~\cite{Kalra}. The coherent control of $^{31}{\rm P}$ nuclear spins is typically achieved by irradiating a silicon sample with resonant RF pulses~\cite{Pla1,Pla3}. The direct effect of such drive pulses on the qubit's environment is negligible, thanks to the large difference in the gyromagnetic ratio between electron and nuclear spins ($\gamma_{e}\gg \gamma_{^{31}{\rm P}} > \gamma_{^{29}{\rm Si}}$). However, the nuclear spin of a $^{31}{\rm P}$ donor indirectly couples to nuclear spins of neighboring $^{29}$Si impurity atoms via the donor qubit's electron. The donor electron enables Fermi contact interaction between the electron spin, and the nuclear spins of the $^{31}{\rm P}$ and $^{29}$Si, resulting in effective nuclear spin-spin coupling. Therefore, coherent rotations of the donor's nuclear spin translate into a net change of the $^{29}$Si nuclear spin bath's Zeeman energy. If the net change is positive, energy is added to the substrate raising its effective temperature. In turn, the thermalization process will increase phonon-mediated spin interactions that may result in a coherence penalty. To avoid this, energy dissipated by qubit control operations must be removed by a refrigerator and may not exceed available cooling capacity. Dilution refrigerators appear to be the most likely solution, although their cooling capacity rapidly diminishes as the operating temperature decreases (e.g., see Fig.~3 in~\cite{Batey}) potentially limiting gate operation rates.   

To evaluate the magnitude of the bath heating effect for silicon qubits, we compute the net Zeeman energy change of $^{29}{\rm Si}$ spins as a function of the number of single-qubit rotations, impurity concentration, and impurity spatial distribution. We find that for a random single rotation about the $X$ axis, the average $^{29}{\rm Si}$ ensemble energy change is negative i.e., the spin bath experiences a cooling effect. However, for a sequence of random single-qubit gates, the average energy change is positive, resulting in bath heating. Further we find that the amount of heating depends on the sequence of single-qubits gates.  As an arbitrary qubit operation may be decomposed in many ways, our results suggest that future quantum computing compilers may need to optimize the selection of gate sequences to reduce thermal effects. 

%\section{The Model}
We begin by introducing a model for the qubit and its environment. The qubit is defined by the nuclear spin ${\bf \hat{I}}_{\rm P}$ of a donor $^{31}{\rm P}$ atom implanted into a silicon substrate. Following recent experiments~\cite{Muhonen}, we will assume that the silicon substrate is purified, and, unless noted otherwise, that the residual concentration of $^{29}{\rm Si}$ is 800 PPM. We will also set the physical qubit volume to 5 nm$^3$, where the scale is set by twice the typical estimate of the Bohr radius of a donor electron~\cite{Saraiva}. The qubit environment then includes nuclear spins of neighboring residual $^{29}{\rm Si}$ atoms ${\bf \hat{I}}_{\rm n}$ (index ${\rm n}$ runs over all lattice sites occupied by $^{29}{\rm Si}$)  and the spin of the donor electron ${\bf \hat{S}}_{e}$. 

When placed in a uniform magnetic field $B_{0}^{z}$ aligned along $Z$ axis, the free Hamiltonian of the qubit reads,
\begin{equation}
\hat{H}_{q} = \omega_{{\rm P}}\hat{I}_{\rm P}^{z},
\end{equation}
where $\omega_{{\rm P}}=\gamma_{\rm P}B^z_0$ and $\gamma_{\rm P}$ is the gyromagnetic ratio for the $^{31}{\rm P}$ nucleus. Similarly, we denote the free Hamiltonian for the donor electron spin, $\hat{H}_{e} = \Omega_{e}\hat{S}_{e}^{z}$ and the neighboring $^{29}{\rm Si}$ nuclear spins,
$\hat{H}_{{\rm Si}} = \sum\limits_{n}\omega_{n}\hat{I}^{z}_{{\rm Si}_{n}}$,
where $\Omega_{e} = \gamma_{e}B^z_0$, $\omega_n = \gamma_{{\rm Si}_{n}}B^z_0$ and $\gamma_{e}$($\gamma_{{\rm Si}_{n}}$) are the electron ($^{29}{\rm Si}$ nucleus) gyromagnetic ratio.

Single-qubit gates are typically implemented as rotations along the $X,Y,Z$ axes by exposing the qubit to time-varying magnetic fields aligned along the axis of rotation. Here, we concentrate on the impact of single-qubit rotations around the $X$ axis on the Zeeman energy of $^{29}{\rm Si}$ spins contained in the physical qubit volume. The Hamiltonian that describes such rotations is 
\begin{equation}\label{Eq:DrivingHamiltonian}
\hat{H}_{d} = \cos(\omega_{d}t)[ \Omega^{x}_{e}\hat{S}^{x}_{e} + \Omega^{x}_{\rm P}\hat{I}^{x}_{\rm P} + \Omega^{x}_{\rm Si}\sum\limits_{n}\hat{I}^{x}_{n}],
\end{equation}
where $B^x_{0}$ and $\omega_{d}$ are the amplitude and the frequency of the AC magnetic field along the $X$ axis, $\Omega^{x}_{e} = \gamma_{e}B^x_0$, $\Omega^{x}_{\rm P} = \gamma_{\rm P}B^x_0$, and $\Omega^{x}_{\rm Si} = \gamma_{\rm Si}B^x_0$. Note that Eq.(\ref{Eq:DrivingHamiltonian}) includes terms corresponding to the effects of the drive field on $^{29}{\rm Si}$ nuclear spins as well as the donor electron spin. As we will show later, the standard ``resonant'' choice of the driving frequency, results in a negligable effect on the electron spin because $\Omega_{e}\gg \omega_{{\rm P}}$.  

Next, we include the effects of the nuclear spin environment. The qubit and $^{29}{\rm Si}$ nuclear spins are coupled to the donor electron spin ${\bf \hat{S}}_{e}$ via the Fermi contact interaction. This coupling, in particular, is the leading cause of  donor electron spin decoherence~\cite{YaoLiuSham,Saikin,Cywinski,Witzel}. In addition, as we will show, this interaction changes the energy of  $^{29}{\rm Si}$ nuclear spins by providing a mechanism for an indirect coupling with $X$-axis qubit rotation pulses. The former represents an example of a well-studied problem, namely, determining the effect of the nuclear spin environment on the qubit. However, the latter poses a less explored question of the thermodynamic effects of qubit operations on the surrounding environment. And while the effects of the environment on the qubit can be largely mitigated e.g., by using dynamic refocusing sequences or higher isotopic purity silicon, removing the effects of qubit control pulses on the spin bath entirely is complex as each qubit will couple to a different random local spin bath. The basic Hamiltonian describing the contact interaction between the donor electron spin and neighboring nuclear $^{31}{\rm P}$ and $^{29}{\rm Si}$ spins is 
\begin{equation}\label{Eq:Hhyperfine}
\hat{H}_{eN} \!= \!\sum\limits_{n\in{\rm Si,P}}\hspace*{-0.2cm}a_n {\bf \hat{S}}_{e}\cdot {\bf \hat{I}}_{n} = \sum\limits_{n\in{\rm Si,P}}\hspace*{-0.2cm} a_n ( \hat{S}^{x}_{e} \hat{I}^{x}_{n} + \hat{S}^{y}_{e} \hat{I}^{y}_{n} + \hat{S}^{z}_{e} \hat{I}^{z}_{n}),
\end{equation}
where hyperfine coupling constants $a_n$ between the donor electron and the n-th nuclear spin are given by
\begin{equation}
a_n = \frac{2\hbar\mu_0}{3}\gamma_e\gamma_n|\Psi({\bf R}_n)|^2.
\end{equation}
Here $\mu_0$ is the vacuum permeability and $\Psi({\bf R}_n)$ is the donor electron wave function at the nucleus location ${\bf R}_n$. Several approaches to modeling the wavefunction $\Psi({\bf R}_n)$ can be found in the literature~\cite{KohnLuttinger,KoillerDonorElectronWF,WellardHollenberg}. For our numerical simulations below we adopt the approximation used to describe the original ENDOR experimental data~\cite{FeherENDOR} with the understanding that its validity is limited to about 20 nuclear shells around the donor~\cite{Hale}. 

In addition to the hyperfine coupling mediated interaction, nuclear spins couple via a direct dipole-dipole interaction. However, this interaction results in pairwise nuclear spin flipping that does not change the overall Zeeman energy of the nuclear spin bath. Also, the dipole-dipole interaction strength is typically weaker than the strength of the hyperfine interaction. Therefore, it is not included in our model.

Finally, besides the $^{29}{\rm Si}$ nuclear spins, $^{31}{\rm P}$ impurities can be introduced during the qubit implantation process. In turn, they can also be polarized by qubit control pulses, leading to additional thermodynamic effects. However, since even in isotopically purified silicon, the number of $^{29}{\rm Si}$ impurities exceeds the number of undesired $^{31}{\rm P}$ donors, we will exclusively focus on estimating the energy deposition to $^{29}{\rm Si}$ environment. 

Putting it all together, we arrive at the total Hamiltonian describing the dynamics of single qubit control pulses coupling to the $^{29}{\rm Si}$ spin bath for a $X$-axis drive field,
\begin{equation}\label{Eq:Hamiltonian}
\hat{H} = \hat{H}_{e} + \hat{H}_{N} + \hat{H}_{d} +\hat{H}_{eN},
\end{equation}
where $\hat{H}_{N} = \hat{H}_{{\rm Si}}+\hat{H}_{q}$.

%\section{Effective Hamiltonian}
In the total Hamiltonian, Eq.(\ref{Eq:Hamiltonian}), three terms may, in principle, change the Zeeman energy of $^{29}{\rm Si}$ nuclei which in turn, can result in an environmental temperature change. These are the the driving term $\hat{H}_{d}$ and the two hyperfine coupling terms $ \sum\limits_{n} a_n \hat{S}^{x}_{e} \hat{I}^{x}_{n}$ and $ \sum\limits_{n} a_n \hat{S}^{y}_{e} \hat{I}^{y}_{n}$. However, because the Zeeman energy of the electron is much larger than the Zeeman energies of phosphorus and silicon nuclei, i.e., $2\hbar\Omega_{e}\gg\{2\hbar\omega_{{\rm P}},2\hbar\omega_{n}\}$, the latter terms can be treated as a small perturbation. It can be shown, by applying perturbation theory, that these terms indeed cancel out to the first and the second order, providing an effective spin-spin coupling that preserves Zeeman energies of the nuclei (a so-called secular term). To see this behaviour we derive the effective Hamiltonian $\hat{H}^{\prime}$ using the method of small rotations~\cite{Klimov}. Accordingly, 
\begin{equation}\label{Eq:rotation}
\hat{H}^{\prime} = U_{r}\hat{H} U^{-1}_{r} 
\end{equation}
where  $U_{r} = \exp(\sum\limits_{n}\alpha_{n}(\sigma^{e}_{+}\sigma^{n}_{-}-\sigma^{e}_{+}\sigma^{n}_{-}))$, $\alpha_n = \frac{a_{n}}{\Delta}$, $\Delta = \Omega_{e}-\omega_{{\rm P}}-\omega_{n}$, and $\sigma^{e}_{\pm}=(\hat{S}^{x}_{e}\pm\hat{S}^{y}_{e})/2$, $\sigma^{n}_{\pm} = (\hat{I}^{x}_{n}\pm\hat{I}^{y}_{n})/2$. For a magnetic field $B^{z}_0=1$ T (a typical experimental value for {\rm Si:P} qubit system) the frequency difference $\Delta$ is on the order of 28 GHz. The hyperfine coupling constants $a_n$ for $^{29}{\rm Si}$ nuclei in a typical qubit volume range between $0.1$ and $10$ MHz, depending on the distance of a $^{29}{\rm Si}$ nucleus from the qubit, while $a_{\rm P}$ for phosphorus is 117 MHz. Hence, for all practical purposes $\alpha_n\ll 1$ and only terms linear in $\alpha_n$ need to remain in Eq.(\ref{Eq:rotation}). Thus $\hat{H}^{\prime}$ becomes,
\begin{equation}
\hat{H}^{\prime} = \hat{H}_{0} + \hat{V},
\end{equation}
where $\hat{H}_{0} = \hat{H}_{e} + \hat{H}_{N}$ and
\begin{eqnarray}
\hat{V} & \approx & \hat{H}_{d} + \sum\limits_{n\in{\rm Si, P}} a_n \hat{S}^{z}_{e} \hat{I}^{z}_{n} + \nonumber \\ 
&  & 2\sum\limits_{n\in{\rm Si, P}}\sum\limits_{m \neq n} \frac{a_{n}a_{m}}{\Delta}\hat{S}^{z}_{e}(\sigma^{n}_{+}\sigma^{m}_{-}+\sigma^{n}_{-}\sigma^{m}_{+}) - \nonumber \\
&  & 2\sum\limits_{n\in{\rm Si, P}}\sum\limits_{m \neq n} \frac{a_{n}a_{m}}{\Delta}(\sigma^{e}_{+}\sigma^{n}_{-}+\sigma^{e}_{-}\sigma^{n}_{+})\hat{I}^{z}_{m} +\nonumber \\
&  & \cos(\omega_{d}t) \sum\limits_{n\in{\rm Si, P}} \frac{\Omega^{e}_{x}a_n}{\Delta}\hat{S}^{z}_{e}\hat{I}^{x}_{n}.
\end{eqnarray}
Here small AC Stark shift terms are omitted because they simply redefine the Zeeman energy of the electron and nuclei. 

Next, we switch to the interaction picture and calculate the effective time-dependent interaction Hamiltonian,
\begin{equation}\label{Eq:IntPicHamiltonian}
\hat{V}(t) = e^{i\hat{H}_{0}t}\hat{V}e^{-i\hat{H}_{0}t}.
\end{equation}
We set the drive field frequency $\omega_{d}$ to be resonant with the qubit transition frequency, i.e., $\omega_{d}\approx\omega_{\rm P}$.  With this drive frequency, qubit rotations around the $X$ axis are facilitated. The corresponding Rabi frequency (the rate of qubit spin flipping along the $Z$ axis) is then determined by the strength of the AC magnetic field $B^{x}_{0}$. Two parameter regimes are possible. First, the ``weak'' driving regime when $B^{x}_{0}\ll B^{z}_{0}$ corresponding to a ``slow'' Rabi flipping with frequencies $<1$ MHz. Second, the ``strong'' driving regime $B^{z}_{0}\gtrsim B^{x}_{0}$ with ``fast'' Rabi flipping frequencies $\ge 10$ MHz. The interaction dynamics $\hat{V}(t)$ is highly dependent on the ratio of the parameters and in the following we consider the two regimes separately.

%\subsection{``Weak'' driving limit}
In the ``weak'' drive scenario $B^{x}_{0}\ll B^{z}_{0}$ which translates into a set of inequalities $\{\Omega^{x}_{e}\ll \Omega_{e}; \Omega^{x}_{\rm P}\ll \omega_{\rm P}; \Omega^{x}_{\rm Si}\ll \omega_{\rm n}\}$. As a result, the rotating wave approximation (RWA) can be applied to the Hamiltonian $\hat{V}(t)$ in Eq.(\ref{Eq:IntPicHamiltonian}).  Here the off-resonant term may be omitted leading to the following effective Hamiltonian:
\begin{eqnarray}\label{Eq:WeakDriving}
\hat{V}_{weak}(t) &\approx& \frac{\Omega^{x}_{\rm P}}{2}\hat{I}^{x}_{\rm P} + \frac{\Omega^{x}_{e}a_{\rm P}}{\Delta}\hat{S}^{z}_{e}\hat{I}^{x}_{\rm P} + \sum\limits_{n\in{\rm Si, P}} a_n \hat{S}^{z}_{e} \hat{I}^{z}_{n} + \nonumber \\ 
&  & 2\sum\limits_{n\in{\rm Si}}\sum\limits_{m \neq n} \frac{a_{n}a_{m}}{\Delta}\hat{S}^{z}_{e}(\sigma^{n}_{+}\sigma^{m}_{-}+\sigma^{n}_{-}\sigma^{m}_{+}). 
\end{eqnarray}
We immediately observe that in the weak drive regime, the Zeeman energy of $^{29}{\rm Si}$ nuclear bath remains unchanged during an arbitrary single qubit rotation gate. Furthermore, the effective Rabi frequency for the qubit now depends on the state of the electron spin. The number of single qubit gates per second is bounded by the effective Rabi frequency. For a drive field $B^{x}_{0}=10^{-3}$ T (with $B^{z}_{0}=1$ T), one can apply $\approx 10^4$ gates per second.

%\subsection{``Strong'' driving limit}
To run a quantum computer at a higher ``clock'' rate, faster single qubit gates are necessary. This can be achieved by applying stronger drive fields while keeping $B^{z}_{0}$ constant. However, once the inequality $B^{x}_{0}\ll B^{z}_{0}$ is violated, the RWA is no longer valid~\cite{Laucht} and the spin dynamics is no longer adequately described by Eq.(\ref{Eq:WeakDriving}). In this case, the effective Hamiltonian Eq.(\ref{Eq:IntPicHamiltonian}) becomes,
\begin{align}
& \hat{V}_{strong}(t)  \approx \sum\limits_{n\in{\rm Si, P}} a_n \hat{S}^{z}_{e} \hat{I}^{z}_{n} \nonumber \\ 
 & +  2\sum\limits_{n\in{\rm Si}}\sum\limits_{m \neq n} \frac{a_{n}a_{m}}{\Delta}\hat{S}^{z}_{e}(\sigma^{n}_{+}\sigma^{m}_{-}+\sigma^{n}_{-}\sigma^{m}_{+})  \nonumber \\
 & +  \,\,\sum\limits_{n\in{\rm Si}} \frac{\Omega^{x}_{e}a_n}{2\Delta}\hat{S}^{z}_{e}[\sigma^{n}_{+}({\rm e}^{2i(\omega_{n}+\omega_{\rm P})t}+{\rm e}^{2i(\omega_{n}-\omega_{\rm P})t})+h.c.]  \nonumber \\
 & +  \frac{\Omega^{x}_{e}a_{\rm P}}{2\Delta}(\hat{S}^{z}_{e}\hat{I}^{x}_{\rm P} + [\hat{S}^{z}_{e}\sigma^{\rm P}_{+}{\rm e}^{4i\omega_{\rm P}t} + h.c.]) \nonumber \\
 & +  \frac{\Omega^{x}_{\rm P}}{2}(\hat{I}^{x}_{\rm P} + [\sigma^{\rm P}_{+}{\rm e}^{4i\omega_{\rm P}t} + h.c.]) \nonumber \\
 & + \frac{\Omega^{x}_{e}}{2}[\sigma^{e}_{+}({\rm e}^{2i(\Omega_{e}+\omega_{\rm P})t}+{\rm e}^{2i(\Omega_{e}-\omega_{\rm P})t})+h.c.] \nonumber \\
 & +  \frac{\Omega^{x}_{\rm Si}}{2}\sum\limits_{n\in {\rm Si}}[\sigma^{n}_{+}({\rm e}^{2i(\omega_{n}+\omega_{\rm P})t}+{\rm e}^{2i(\omega_{n}-\omega_{\rm P})t})+h.c.],\label{Eq:EffectiveStrong}
\end{align}
where we have omitted terms of the order $\frac{a_{n}a_{\rm P}}{\Delta|\omega_{\rm P}-\omega_{n}|}$ and $\frac{a_{n}a_{\rm P}}{\Delta^2}$ $\approx (\frac{a_{n}}{\Delta})^2\ll 1$. Note that now there are non-vanishing terms that act to periodically drive the $^{29}{\rm Si}$ nuclear spins. This will result in a Zeeman energy change of the nuclear spin bath. To evaluate the effect of these terms we solve the Schr{\"o}dinger equation for the qubit-electron-nuclear spin wave function using time-dependent perturbation theory. First, we eliminate the time-independent terms in Eq.(\ref{Eq:EffectiveStrong}) by applying an additional unitary rotation, giving
\begin{equation}
\hat{V}^{\prime}_{strong}(t) =  e^{i \hat{H}_{ZZ}t}\hat{V}_{strong}(t)e^{-i \hat{H}_{ZZ}t} 
\end{equation} 
where $\hat{H}_{ZZ} = \sum\limits_{n\in{\rm Si, P}} a_n \hat{S}^{z}_{e} \hat{I}^{z}_{n}$. After some algebra we obtain,
\begin{eqnarray}\label{Eq:EffectiveStrongFinal}
\hat{V}^{\prime}_{strong}(t) &\approx& \sum\limits_{n\in{\rm Si,P}} (\frac{\Omega^{x}_{e}a_n}{2\Delta}\hat{S}^{z}_{e}+\frac{\Omega^{x}_{n}}{2}1\!\!1)[({\rm e}^{2i(\omega_{n}+\omega_{\rm P}+a_n\hat{S}^{z}_{e})t}  \nonumber \\
& + & {\rm e}^{2i(\omega_{n}-\omega_{\rm P}+a_n\hat{S}^{z}_{e})t})\sigma^{n}_{+}+h.c.] \nonumber \\
& + & \frac{\Omega^{x}_{e}}{2}[({\rm e}^{2i(\Omega_{e}+\omega_{\rm P}+\sum\limits_{n\in{\rm Si}} a_n\hat{I}^{z}_{n})t} \nonumber \\
& + & {\rm e}^{2i(\Omega_{e}-\omega_{\rm P}+\sum\limits_{n\in{\rm Si}} a_n\hat{I}^{z}_{n})t})\sigma^{e}_{+}+h.c.]. 
\end{eqnarray}
In the last expression, we have omitted terms of the order $\frac{a_{n}a_{n}}{\Delta|a_{n}-a_{m}|}\ll 1$. 

Next, we calculate the propagator $U(t_0,t)$ to the first order of the Dyson series using the time dependent Hamiltonian in Eq.(\ref{Eq:EffectiveStrongFinal}),
\begin{equation}\label{Eq:Propagator}
U(t_0,t) \approx 1\!\!1 - i\int\limits_{t_0}^{t}d\tau\hat{V}^{\prime}_{strong}(\tau).
\end{equation}
%It is straightforward to see that the integration in Eq.(\ref{Eq:Propagator}) can be performed %analytically. We notice that,
%\begin{equation}\label{Eq:OpIntegral}
%\int\limits_{t_0}^{t}d\tau \exp(i\hat{O}\tau) = i\hat{O}^{-1}(\exp(i\hat{O}t)-\exp(i\hat{O}t_0)).
%\end{equation}
%If $\hat{O}^{-1}$ exists, then integrals of the operators are expressible as $\exp(i(\alpha 1\!\!1 + \beta\hat{S}^{z}_{e}))$, which is analytically integratable. After some algebra, the inverse operator may be written as
%\begin{equation}\label{Eq:OpInverse}
%i(\alpha 1\!\!1 + \beta\hat{S}^{z}_{e})^{-1}=\frac{i}{\beta^2-\alpha^2}(\alpha 1\!\!1 - \beta\hat{S}^{z}_{e}).
%\end{equation} 
Assuming that the qubit and its spin bath are initially in a separable state $|\psi_{q+b}(t_{0})\rangle=|\psi_{q}(t_{0})\rangle\otimes|\psi_{b}(t_{0})\rangle$, the evolution of the state for the combined qubit-spin-bath system in the interaction picture is given by,
\begin{equation}
|\psi_{q+b}(t)\rangle \approx U(t_0,t)U_{r}^{\dagger}|\psi_{q+b}(t_{0})\rangle,
\end{equation}
for sufficiently short time intervals $\Delta t = t-t_{0}$. The change in the Zeeman energy of the $^{29}{\rm Si}$ nuclear spin bath can then be evaluated:
\begin{equation}\label{Eq:deltaE}
\hspace*{-0.3cm}\frac{\Delta E_{{\rm Si}}(t)}{\hbar}=\langle \psi_{q+b}(0)|U_{r}\left(\hat{H}_{{\rm Si}}(t)-\hat{H}_{{\rm Si}}\right)U_{r}^{\dagger}|\psi_{q+b}(0)\rangle, 
\end{equation}
where we have defined $\hat{H}_{{\rm Si}}(t) = U^{\dagger}(0,t)U_{I}(t)\hat{H}_{{\rm Si}}U^{\dagger}_{I}(t)U(0,t)$ and $U_{I}(t)$ is a unitary rotation into the interaction picture. In what follows we numerically evaluate $\Delta E_{{\rm Si}}(t)$ as a function of the spatial distribution and the concentration of $^{29}{\rm Si}$ nuclear spins as well as a function of single-qubit rotations.

\begin{figure}[t!]
    \centering
    \includegraphics[width=0.5\textwidth]{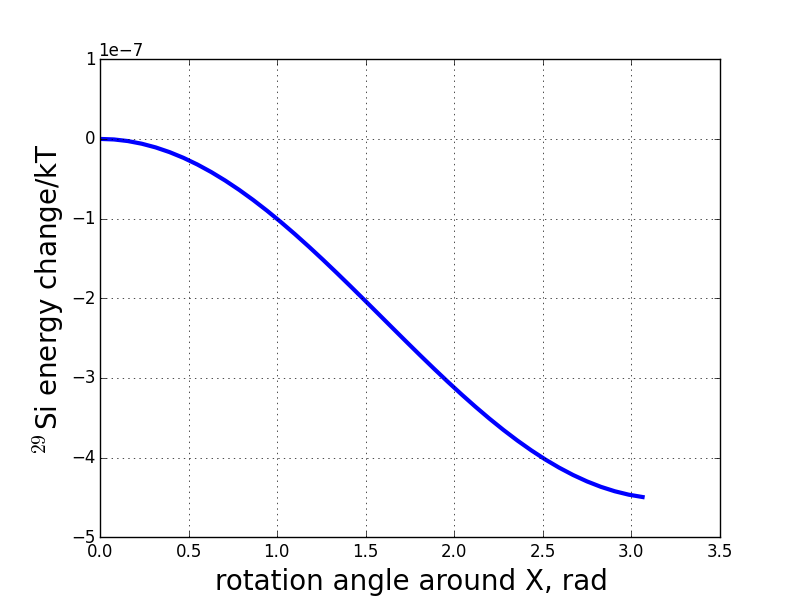}
    \caption{(Color Online) Average Zeeman energy change of the $^{29}{\rm Si}$ nuclear spin (normalized to $kT$) as a function of the single-qubit rotation angle around $X$ axis.}
    \label{fig:fig1}
\end{figure}

%\section{Simulation Results}
%\subsection{Single Rotation}
Our first set of simulations is aimed at studying the spin bath effect from applying a {\it single} gate to the qubit. For the sake of concreteness, we choose a rotation around the $X$ axis by an angle $\phi$ ($\phi\in [0,\pi]$). First, we set the $^{29}{\rm Si}$ concentration to 800 PPM and generate 200 random spatial $^{29}{\rm Si}$ nuclear spin distributions.  We assume that the $^{31}{\rm P}$ nucleus is located at the center of a 5 nm$^3$ cube. The spatial distribution sampling is performed to account for the silicon substrate variability across a multi-qubit quantum device that is inevitable due to the random nature of $^{29}{\rm Si}$ inclusions. We further assume that the qubit's initial state is always spin down $|\!\!\Downarrow \rangle$ and the donor electron is initialized into the spin-up state $|\!\!\uparrow \rangle$. Then for each spatial spin bath configuration, and every possible nuclear spin-bath state ranging from $|\!\!\Downarrow,\cdots,\Downarrow \rangle$ to $|\!\!\Uparrow,\cdots,\Uparrow \rangle$ (for a total of $2^{N}$ states where $N$ is the number of $^{29}{\rm Si}$ nuclear spins per qubit volume), we compute $\Delta E_{{\rm Si}}(t)$ ($t\in [0,\frac{\pi}{\Omega^{x}_{{\rm P}}}]$) using Eq.(\ref{Eq:deltaE}). Finally, for each rotation angle $\phi$ or, equivalently, time $t$, we compute the mean spin bath energy change $\langle\Delta E_{{\rm Si}}(t)\rangle$ by averaging over all spin bath spatial configurations and states. The spin state averaging is performed assuming that the spin-bath states are distributed according to the Boltzmann distribution. In Fig.~\ref{fig:fig1}, we plot $\langle\Delta E_{{\rm Si}}(t)\rangle$ normalized to the thermal energy $kT$ (assuming $B^{z}_{0} = 1~{\rm T}$, $B^{x}_{0} = 100 ~{\rm mT}$ and the ambient temperature is $250~{\rm mK}$). We observe that driving the qubit coherently from  $|\!\!\Downarrow \rangle$ to  $|\!\!\Uparrow \rangle$ on average decreases the Zeeman energy of the $^{29}{\rm Si}$ spin bath, effectively lowering the temperature of the bath. As one would intuitively expect, the largest energy change happens when the qubit state is flipped completely around the $X$-axis via a $\pi$ rotation. 

\begin{figure}[t!]
    \centering
    \includegraphics[width=0.5\textwidth]{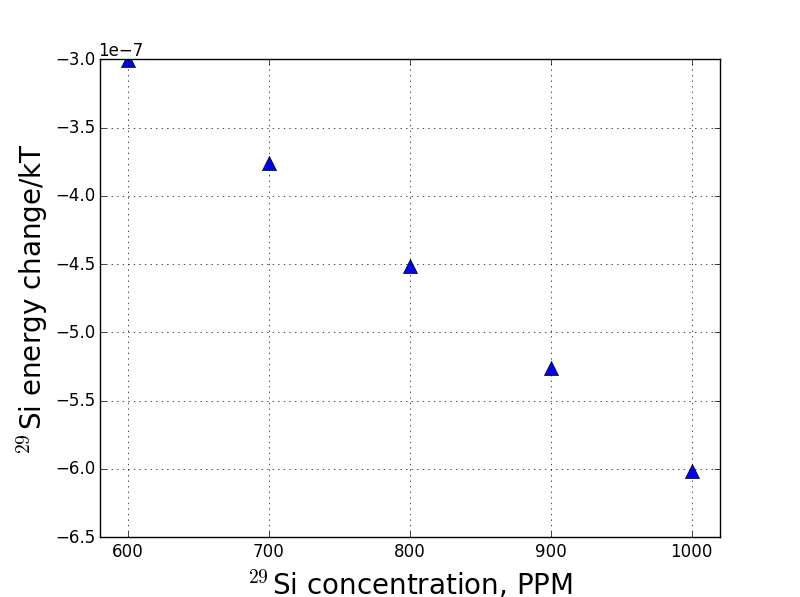}
    \caption{(Color Online) Average Zeeman energy change of the $^{29}{\rm Si}$ nuclear spin (normalized to $kT$) for a single $\pi$ rotation around $X$ axis as a function of the $^{29}{\rm Si}$ concentration.}
    \label{fig:fig2}
\end{figure}

Next, we examine how the observed cooling effect depends on the concentration of $^{29}{\rm Si}$ nuclear spins. We fix the qubit rotation angle around $X$ to $\pi$. Then for a range of $^{29}{\rm Si}$ concentrations we repeat the simulation steps outlined in the preceding paragraph and compute  $\langle\Delta E_{{\rm Si}}(\frac{\pi}{\Omega^{x}_{{\rm P}}})\rangle$. In Fig.~\ref{fig:fig2} we plot the resulting concentration dependence. We immediately notice that the bath cooling effect increases linearly with the concentration of $^{29}{\rm Si}$ nuclear spins. At the same time, higher concentrations of $^{29}{\rm Si}$ affect qubit's coherence. Keeping the impurity concentration low minimizes effects of qubit controls on the environment as well as the effects of the environment on qubit coherence.   

\begin{figure}
    \centering
    \includegraphics[width=0.5\textwidth]{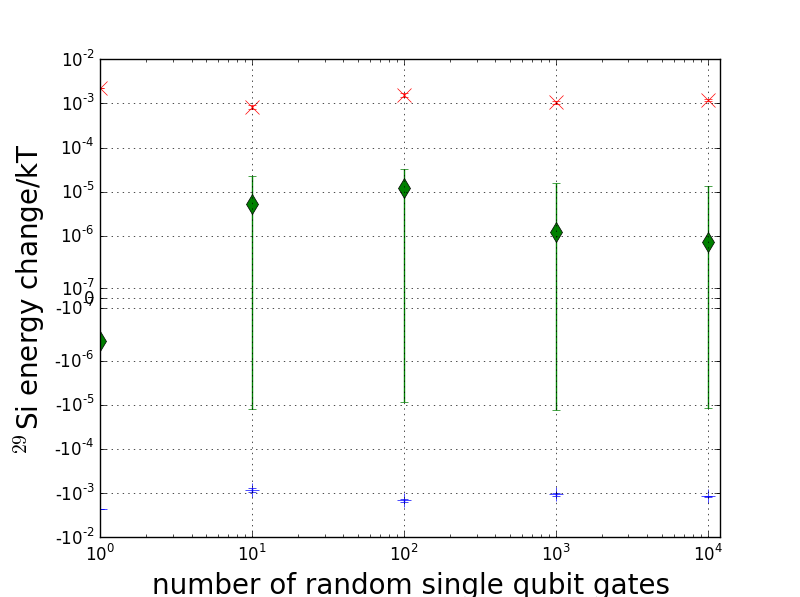}
    \caption{(Color Online) Average (green diamonds), maximum (red $\times$), and minimum (blue +) Zeeman energy change of the $^{29}{\rm Si}$ nuclear spin (normalized to $kT$) as a function of the single-qubit rotation sequence length.}
    \label{fig:fig3}
\end{figure}

%\subsection{Random Rotation Sequences}
In the preceding discussion, we describe the thermodynamic effects from a single qubit rotation. However, any practically significant quantum algorithm will require the implementation of  thousands of single qubit gates run in a sequence. Because, the single-qubit gates in the sequence are interleaved with periods of free spin evolution, we expect that the net thermodynamic effect on the spin bath will be different and will depend on the length of the sequence. To test this hypothesis we generated 120 random $^{29}{\rm Si}$ spatial spin distributions for a 800 PPM concentration of $^{29}{\rm Si}$. For every spatial distribution we initialize the qubit, the donor electron spin, and $^{29}{\rm Si}$ nuclear spin bath state into the $|\!\!\downarrow \rangle$, $|\!\!\uparrow \rangle$, and one of the Boltzmann ensemble states, respectively. Then we apply a sequence of $10^{4}$ random qubit rotations around the $X$ axis ($\{\phi_{1},\cdots,\phi_{10000}\}, \phi_{i}\in [0,2\pi]$). After each rotation, the $^{29}{\rm Si}$ nuclear spin bath energy change with respect to its initial value (prior to all rotations) is recorded. We allow the qubit-environment system to evolve freely in between the rotations for the time duration $\tau = \frac{\pi}{\Omega^{x}_{{\rm P}}}$. Lastly, for the rotation sequences of the length $10^{i}, i=0,\cdots,4$, we compute the change in the spin-bath energy $\langle\Delta E_{{\rm Si}}\rangle_{i}$, averaged over the spatial and Boltzmann spin distributions. We plot the results in Fig.~\ref{fig:fig3}. The red  $\times$ (blue +) represent the maximum (minimum) spin bath energy change averaged over all spatial configurations. The green diamonds are the ``doubly'' averaged (over the spatial and Boltzmann distributions) energy change $\langle\Delta E_{{\rm Si}}\rangle_{i}$. The vertical bars show the variance over 120 spatial spin location configurations. Note that for a gate sequence of length 1, the average energy change is negative which is consistent with our previous results. However, for longer gate sequences, the average energy change is positive, i.e., on average the spin bath temperature increases. This effect strongly depends on the spin-bath spatial locations: certain configurations will always experience cooling as evidenced by the energy distribution variance.  

%\section{Summary}
We have studied the effects of single-qubit operations on the $^{29}{\rm Si}$ nuclear spin bath in Si:P based quantum computing. We developed a Hamiltonian model that describes an effective coupling mechanism between qubit microwave control pulses and the nuclear spin bath. Using a combination of analytical and numerical tools, we have simulated the behavior of the nuclear spin bath as a function of applied single-qubit rotations. We showed that when only a single rotation is applied to the qubit, on average, the nuclear bath energy decreases, equivalent to cooling. However, for arbitrary sequences of rotations, the average effect results in heating. Although the magnitude of the heating/cooling effects is small for a single qubit, for a fully error-corrected quantum computer with $10^{7}$ qubits, the cumulative effect is non-negligible. While our analysis shows there is parasitic coupling of the single-qubit control fields to the nuclear spin bath, it should be manageable for near-term silicon-based qubit technology.   

%\begin{acknowledgments}
{\it Acknowledgments.} We thank Fahd A. Mohiyaddin and Ryan S. Bennink for useful discussions. This work was performed at Oak Ridge National Laboratory, operated by UT-Battelle for the U.S. Department of Energy under contract no. DE-AC05-00OR22725. Research sponsored by the Laboratory Directed Research and Development Program of Oak Ridge National Laboratory,
managed by UT-Battelle, LLC, for the U. S. Department of Energy.
%\end{acknowledgments}

%\onecolumngrid
%\begin{figure}
%    \centering
%    \includegraphics[width=1.0\textwidth]{GateSeguences10_10000_no_err_bars.png}
%    \caption{(Color Online)}
%    \label{fig:fig4}
%\end{figure}
%\twocolumngrid

%


\begin{thebibliography}{28}%
\makeatletter
\providecommand \@ifxundefined [1]{%
 \@ifx{#1\undefined}
}%
\providecommand \@ifnum [1]{%
 \ifnum #1\expandafter \@firstoftwo
 \else \expandafter \@secondoftwo
 \fi
}%
\providecommand \@ifx [1]{%
 \ifx #1\expandafter \@firstoftwo
 \else \expandafter \@secondoftwo
 \fi
}%
\providecommand \natexlab [1]{#1}%
\providecommand \enquote  [1]{``#1''}%
\providecommand \bibnamefont  [1]{#1}%
\providecommand \bibfnamefont [1]{#1}%
\providecommand \citenamefont [1]{#1}%
\providecommand \href@noop [0]{\@secondoftwo}%
\providecommand \href [0]{\begingroup \@sanitize@url \@href}%
\providecommand \@href[1]{\@@startlink{#1}\@@href}%
\providecommand \@@href[1]{\endgroup#1\@@endlink}%
\providecommand \@sanitize@url [0]{\catcode `\\12\catcode `\$12\catcode
  `\&12\catcode `\#12\catcode `\^12\catcode `\_12\catcode `\%12\relax}%
\providecommand \@@startlink[1]{}%
\providecommand \@@endlink[0]{}%
\providecommand \url  [0]{\begingroup\@sanitize@url \@url }%
\providecommand \@url [1]{\endgroup\@href {#1}{\urlprefix }}%
\providecommand \urlprefix  [0]{URL }%
\providecommand \Eprint [0]{\href }%
\providecommand \doibase [0]{http://dx.doi.org/}%
\providecommand \selectlanguage [0]{\@gobble}%
\providecommand \bibinfo  [0]{\@secondoftwo}%
\providecommand \bibfield  [0]{\@secondoftwo}%
\providecommand \translation [1]{[#1]}%
\providecommand \BibitemOpen [0]{}%
\providecommand \bibitemStop [0]{}%
\providecommand \bibitemNoStop [0]{.\EOS\space}%
\providecommand \EOS [0]{\spacefactor3000\relax}%
\providecommand \BibitemShut  [1]{\csname bibitem#1\endcsname}%
\let\auto@bib@innerbib\@empty
%</preamble>
\bibitem [{\citenamefont {Ladd}\ \emph {et~al.}(2010)\citenamefont {Ladd},
  \citenamefont {Jelezko}, \citenamefont {Laflamme}, \citenamefont {Nakamura},
  \citenamefont {Monroe},\ and\ \citenamefont {O'Brien}}]{Ladd}%
  \BibitemOpen
  \bibfield  {author} {\bibinfo {author} {\bibfnamefont {T.~D.}\ \bibnamefont
  {Ladd}}, \bibinfo {author} {\bibfnamefont {F.}~\bibnamefont {Jelezko}},
  \bibinfo {author} {\bibfnamefont {R.}~\bibnamefont {Laflamme}}, \bibinfo
  {author} {\bibfnamefont {Y.}~\bibnamefont {Nakamura}}, \bibinfo {author}
  {\bibfnamefont {C.}~\bibnamefont {Monroe}}, \ and\ \bibinfo {author}
  {\bibfnamefont {J.~L.}\ \bibnamefont {O'Brien}},\ }\href
  {http://dx.doi.org/10.1038/nature08812} {\bibfield  {journal} {\bibinfo
  {journal} {Nature}\ }\textbf {\bibinfo {volume} {464}},\ \bibinfo {pages}
  {45} (\bibinfo {year} {2010})}\BibitemShut {NoStop}%
\bibitem [{\citenamefont {Suter}\ and\ \citenamefont
  {\'Alvarez}(2016)}]{RevModPhys.88.041001}%
  \BibitemOpen
  \bibfield  {author} {\bibinfo {author} {\bibfnamefont {D.}~\bibnamefont
  {Suter}}\ and\ \bibinfo {author} {\bibfnamefont {G.~A.}\ \bibnamefont
  {\'Alvarez}},\ }\href {\doibase 10.1103/RevModPhys.88.041001} {\bibfield
  {journal} {\bibinfo  {journal} {Rev. Mod. Phys.}\ }\textbf {\bibinfo {volume}
  {88}},\ \bibinfo {pages} {041001} (\bibinfo {year} {2016})}\BibitemShut
  {NoStop}%
\bibitem [{\citenamefont {Lidar}(2014)}]{lidar14}%
  \BibitemOpen
  \bibfield  {author} {\bibinfo {author} {\bibfnamefont {D.~A.}\ \bibnamefont
  {Lidar}},\ }\enquote {\bibinfo {title} {Review of decoherence-free subspaces,
  noiseless subsystems, and dynamical decoupling},}\ in\ \href {\doibase
  10.1002/9781118742631.ch11} {\emph {\bibinfo {booktitle} {Quantum Information
  and Computation for Chemistry}}}\ (\bibinfo  {publisher} {John Wiley \& Sons,
  Inc.},\ \bibinfo {year} {2014})\ pp.\ \bibinfo {pages} {295--354}\BibitemShut
  {NoStop}%
\bibitem [{\citenamefont {Gaitan}(2007)}]{Gaitan:2007:QEC:1554882}%
  \BibitemOpen
  \bibfield  {author} {\bibinfo {author} {\bibfnamefont {F.}~\bibnamefont
  {Gaitan}},\ }\href@noop {} {\emph {\bibinfo {title} {Quantum Error Correction
  and Fault Tolerant Quantum Computing}}}\ (\bibinfo  {publisher} {CRC Press,
  Inc.},\ \bibinfo {address} {Boca Raton, FL, USA},\ \bibinfo {year}
  {2007})\BibitemShut {NoStop}%
\bibitem [{\citenamefont {Lidar}\ and\ \citenamefont
  {Brun}(2013)}]{lidar2013quantum}%
  \BibitemOpen
  \bibfield  {author} {\bibinfo {author} {\bibfnamefont {D.}~\bibnamefont
  {Lidar}}\ and\ \bibinfo {author} {\bibfnamefont {T.}~\bibnamefont {Brun}},\
  }\href {https://books.google.com/books?id=bM5KngEACAAJ} {\emph {\bibinfo
  {title} {Quantum Error Correction}}}\ (\bibinfo  {publisher} {Cambridge
  University Press},\ \bibinfo {year} {2013})\BibitemShut {NoStop}%
\bibitem [{\citenamefont {Kane}(1998)}]{Kane}%
  \BibitemOpen
  \bibfield  {author} {\bibinfo {author} {\bibfnamefont {B.~E.}\ \bibnamefont
  {Kane}},\ }\href@noop {} {\bibfield  {journal} {\bibinfo  {journal} {Nature}\
  }\textbf {\bibinfo {volume} {393}},\ \bibinfo {pages} {133} (\bibinfo {year}
  {1998})}\BibitemShut {NoStop}%
\bibitem [{\citenamefont {Pla}\ \emph {et~al.}(2012)\citenamefont {Pla},
  \citenamefont {Tan}, \citenamefont {Dehollain}, \citenamefont {Lim},
  \citenamefont {Morton}, \citenamefont {Jamieson}, \citenamefont {Dzurak},\
  and\ \citenamefont {Morello}}]{Pla2}%
  \BibitemOpen
  \bibfield  {author} {\bibinfo {author} {\bibfnamefont {J.~J.}\ \bibnamefont
  {Pla}}, \bibinfo {author} {\bibfnamefont {K.~Y.}\ \bibnamefont {Tan}},
  \bibinfo {author} {\bibfnamefont {J.~P.}\ \bibnamefont {Dehollain}}, \bibinfo
  {author} {\bibfnamefont {W.~H.}\ \bibnamefont {Lim}}, \bibinfo {author}
  {\bibfnamefont {J.~J.}\ \bibnamefont {Morton}}, \bibinfo {author}
  {\bibfnamefont {D.~N.}\ \bibnamefont {Jamieson}}, \bibinfo {author}
  {\bibfnamefont {A.~S.}\ \bibnamefont {Dzurak}}, \ and\ \bibinfo {author}
  {\bibfnamefont {A.}~\bibnamefont {Morello}},\ }\href@noop {} {\bibfield
  {journal} {\bibinfo  {journal} {Nature}\ }\textbf {\bibinfo {volume} {489}},\
  \bibinfo {pages} {541} (\bibinfo {year} {2012})}\BibitemShut {NoStop}%
\bibitem [{\citenamefont {Pla}\ \emph {et~al.}(2013)\citenamefont {Pla},
  \citenamefont {Tan}, \citenamefont {Dehollain}, \citenamefont {Lim},
  \citenamefont {Morton}, \citenamefont {Zwanenburg}, \citenamefont {Jamieson},
  \citenamefont {Dzurak},\ and\ \citenamefont {Morello}}]{Pla1}%
  \BibitemOpen
  \bibfield  {author} {\bibinfo {author} {\bibfnamefont {J.~J.}\ \bibnamefont
  {Pla}}, \bibinfo {author} {\bibfnamefont {K.~Y.}\ \bibnamefont {Tan}},
  \bibinfo {author} {\bibfnamefont {J.~P.}\ \bibnamefont {Dehollain}}, \bibinfo
  {author} {\bibfnamefont {W.~H.}\ \bibnamefont {Lim}}, \bibinfo {author}
  {\bibfnamefont {J.~J.}\ \bibnamefont {Morton}}, \bibinfo {author}
  {\bibfnamefont {F.~A.}\ \bibnamefont {Zwanenburg}}, \bibinfo {author}
  {\bibfnamefont {D.~N.}\ \bibnamefont {Jamieson}}, \bibinfo {author}
  {\bibfnamefont {A.~S.}\ \bibnamefont {Dzurak}}, \ and\ \bibinfo {author}
  {\bibfnamefont {A.}~\bibnamefont {Morello}},\ }\href@noop {} {\bibfield
  {journal} {\bibinfo  {journal} {Nature}\ }\textbf {\bibinfo {volume} {496}},\
  \bibinfo {pages} {334} (\bibinfo {year} {2013})}\BibitemShut {NoStop}%
\bibitem [{\citenamefont {Muhonen}\ \emph {et~al.}(2014)\citenamefont
  {Muhonen}, \citenamefont {Dehollain}, \citenamefont {Laucht}, \citenamefont
  {Hudson}, \citenamefont {Kalra}, \citenamefont {Sekiguchi}, \citenamefont
  {Itoh}, \citenamefont {Jamieson}, \citenamefont {McCallum}, \citenamefont
  {Dzurak} \emph {et~al.}}]{Muhonen}%
  \BibitemOpen
  \bibfield  {author} {\bibinfo {author} {\bibfnamefont {J.~T.}\ \bibnamefont
  {Muhonen}}, \bibinfo {author} {\bibfnamefont {J.~P.}\ \bibnamefont
  {Dehollain}}, \bibinfo {author} {\bibfnamefont {A.}~\bibnamefont {Laucht}},
  \bibinfo {author} {\bibfnamefont {F.~E.}\ \bibnamefont {Hudson}}, \bibinfo
  {author} {\bibfnamefont {R.}~\bibnamefont {Kalra}}, \bibinfo {author}
  {\bibfnamefont {T.}~\bibnamefont {Sekiguchi}}, \bibinfo {author}
  {\bibfnamefont {K.~M.}\ \bibnamefont {Itoh}}, \bibinfo {author}
  {\bibfnamefont {D.~N.}\ \bibnamefont {Jamieson}}, \bibinfo {author}
  {\bibfnamefont {J.~C.}\ \bibnamefont {McCallum}}, \bibinfo {author}
  {\bibfnamefont {A.~S.}\ \bibnamefont {Dzurak}},  \emph {et~al.},\ }\href@noop
  {} {\bibfield  {journal} {\bibinfo  {journal} {Nature nanotechnology}\
  }\textbf {\bibinfo {volume} {9}},\ \bibinfo {pages} {986} (\bibinfo {year}
  {2014})}\BibitemShut {NoStop}%
\bibitem [{\citenamefont {Pla}\ \emph {et~al.}(2014)\citenamefont {Pla},
  \citenamefont {Mohiyaddin}, \citenamefont {Tan}, \citenamefont {Dehollain},
  \citenamefont {Rahman}, \citenamefont {Klimeck}, \citenamefont {Jamieson},
  \citenamefont {Dzurak},\ and\ \citenamefont {Morello}}]{Pla3}%
  \BibitemOpen
  \bibfield  {author} {\bibinfo {author} {\bibfnamefont {J.~J.}\ \bibnamefont
  {Pla}}, \bibinfo {author} {\bibfnamefont {F.~A.}\ \bibnamefont {Mohiyaddin}},
  \bibinfo {author} {\bibfnamefont {K.~Y.}\ \bibnamefont {Tan}}, \bibinfo
  {author} {\bibfnamefont {J.~P.}\ \bibnamefont {Dehollain}}, \bibinfo {author}
  {\bibfnamefont {R.}~\bibnamefont {Rahman}}, \bibinfo {author} {\bibfnamefont
  {G.}~\bibnamefont {Klimeck}}, \bibinfo {author} {\bibfnamefont {D.~N.}\
  \bibnamefont {Jamieson}}, \bibinfo {author} {\bibfnamefont {A.~S.}\
  \bibnamefont {Dzurak}}, \ and\ \bibinfo {author} {\bibfnamefont
  {A.}~\bibnamefont {Morello}},\ }\href {\doibase
  10.1103/PhysRevLett.113.246801} {\bibfield  {journal} {\bibinfo  {journal}
  {Phys. Rev. Lett.}\ }\textbf {\bibinfo {volume} {113}},\ \bibinfo {pages}
  {246801} (\bibinfo {year} {2014})}\BibitemShut {NoStop}%
\bibitem [{\citenamefont {B{\"u}ch}\ \emph {et~al.}(2013)\citenamefont
  {B{\"u}ch}, \citenamefont {Mahapatra}, \citenamefont {Rahman}, \citenamefont
  {Morello},\ and\ \citenamefont {Simmons}}]{Buch2013}%
  \BibitemOpen
  \bibfield  {author} {\bibinfo {author} {\bibfnamefont {H.}~\bibnamefont
  {B{\"u}ch}}, \bibinfo {author} {\bibfnamefont {S.}~\bibnamefont {Mahapatra}},
  \bibinfo {author} {\bibfnamefont {R.}~\bibnamefont {Rahman}}, \bibinfo
  {author} {\bibfnamefont {A.}~\bibnamefont {Morello}}, \ and\ \bibinfo
  {author} {\bibfnamefont {M.}~\bibnamefont {Simmons}},\ }\href@noop {}
  {\bibfield  {journal} {\bibinfo  {journal} {Nature communications}\ }\textbf
  {\bibinfo {volume} {4}},\ \bibinfo {pages} {2017} (\bibinfo {year}
  {2013})}\BibitemShut {NoStop}%
\bibitem [{\citenamefont {Becker}\ \emph {et~al.}(2010)\citenamefont {Becker},
  \citenamefont {Pohl}, \citenamefont {Riemann},\ and\ \citenamefont
  {Abrosimov}}]{Becker}%
  \BibitemOpen
  \bibfield  {author} {\bibinfo {author} {\bibfnamefont {P.}~\bibnamefont
  {Becker}}, \bibinfo {author} {\bibfnamefont {H.-J.}\ \bibnamefont {Pohl}},
  \bibinfo {author} {\bibfnamefont {H.}~\bibnamefont {Riemann}}, \ and\
  \bibinfo {author} {\bibfnamefont {N.}~\bibnamefont {Abrosimov}},\ }\href
  {\doibase 10.1002/pssa.200925148} {\bibfield  {journal} {\bibinfo  {journal}
  {physica status solidi (a)}\ }\textbf {\bibinfo {volume} {207}},\ \bibinfo
  {pages} {49} (\bibinfo {year} {2010})}\BibitemShut {NoStop}%
\bibitem [{\citenamefont {Yao}\ \emph {et~al.}(2006)\citenamefont {Yao},
  \citenamefont {Liu},\ and\ \citenamefont {Sham}}]{YaoLiuSham}%
  \BibitemOpen
  \bibfield  {author} {\bibinfo {author} {\bibfnamefont {W.}~\bibnamefont
  {Yao}}, \bibinfo {author} {\bibfnamefont {R.-B.}\ \bibnamefont {Liu}}, \ and\
  \bibinfo {author} {\bibfnamefont {L.~J.}\ \bibnamefont {Sham}},\ }\href
  {\doibase 10.1103/PhysRevB.74.195301} {\bibfield  {journal} {\bibinfo
  {journal} {Phys. Rev. B}\ }\textbf {\bibinfo {volume} {74}},\ \bibinfo
  {pages} {195301} (\bibinfo {year} {2006})}\BibitemShut {NoStop}%
\bibitem [{\citenamefont {Saikin}\ \emph {et~al.}(2007)\citenamefont {Saikin},
  \citenamefont {Yao},\ and\ \citenamefont {Sham}}]{Saikin}%
  \BibitemOpen
  \bibfield  {author} {\bibinfo {author} {\bibfnamefont {S.~K.}\ \bibnamefont
  {Saikin}}, \bibinfo {author} {\bibfnamefont {W.}~\bibnamefont {Yao}}, \ and\
  \bibinfo {author} {\bibfnamefont {L.~J.}\ \bibnamefont {Sham}},\ }\href
  {\doibase 10.1103/PhysRevB.75.125314} {\bibfield  {journal} {\bibinfo
  {journal} {Phys. Rev. B}\ }\textbf {\bibinfo {volume} {75}},\ \bibinfo
  {pages} {125314} (\bibinfo {year} {2007})}\BibitemShut {NoStop}%
\bibitem [{\citenamefont {Cywi{\'n}ski}\ \emph {et~al.}(2009)\citenamefont
  {Cywi{\'n}ski}, \citenamefont {Witzel},\ and\ \citenamefont
  {Sarma}}]{Cywinski}%
  \BibitemOpen
  \bibfield  {author} {\bibinfo {author} {\bibfnamefont {{\L}.}~\bibnamefont
  {Cywi{\'n}ski}}, \bibinfo {author} {\bibfnamefont {W.~M.}\ \bibnamefont
  {Witzel}}, \ and\ \bibinfo {author} {\bibfnamefont {S.~D.}\ \bibnamefont
  {Sarma}},\ }\href@noop {} {\bibfield  {journal} {\bibinfo  {journal} {Phys.
  Rev. Lett.}\ }\textbf {\bibinfo {volume} {102}},\ \bibinfo {pages} {057601}
  (\bibinfo {year} {2009})}\BibitemShut {NoStop}%
\bibitem [{\citenamefont {Tyryshkin}\ \emph {et~al.}(2012)\citenamefont
  {Tyryshkin}, \citenamefont {Tojo}, \citenamefont {Morton}, \citenamefont
  {Riemann}, \citenamefont {Abrosimov}, \citenamefont {Becker}, \citenamefont
  {Pohl}, \citenamefont {Schenkel}, \citenamefont {Thewalt}, \citenamefont
  {Itoh} \emph {et~al.}}]{Tyryshkin}%
  \BibitemOpen
  \bibfield  {author} {\bibinfo {author} {\bibfnamefont {A.~M.}\ \bibnamefont
  {Tyryshkin}}, \bibinfo {author} {\bibfnamefont {S.}~\bibnamefont {Tojo}},
  \bibinfo {author} {\bibfnamefont {J.~J.}\ \bibnamefont {Morton}}, \bibinfo
  {author} {\bibfnamefont {H.}~\bibnamefont {Riemann}}, \bibinfo {author}
  {\bibfnamefont {N.~V.}\ \bibnamefont {Abrosimov}}, \bibinfo {author}
  {\bibfnamefont {P.}~\bibnamefont {Becker}}, \bibinfo {author} {\bibfnamefont
  {H.-J.}\ \bibnamefont {Pohl}}, \bibinfo {author} {\bibfnamefont
  {T.}~\bibnamefont {Schenkel}}, \bibinfo {author} {\bibfnamefont {M.~L.}\
  \bibnamefont {Thewalt}}, \bibinfo {author} {\bibfnamefont {K.~M.}\
  \bibnamefont {Itoh}},  \emph {et~al.},\ }\href@noop {} {\bibfield  {journal}
  {\bibinfo  {journal} {Nature materials}\ }\textbf {\bibinfo {volume} {11}},\
  \bibinfo {pages} {143} (\bibinfo {year} {2012})}\BibitemShut {NoStop}%
\bibitem [{\citenamefont {Hill}\ \emph {et~al.}(2015)\citenamefont {Hill},
  \citenamefont {Peretz}, \citenamefont {Hile}, \citenamefont {House},
  \citenamefont {Fuechsle}, \citenamefont {Rogge}, \citenamefont {Simmons},\
  and\ \citenamefont {Hollenberg}}]{Hill}%
  \BibitemOpen
  \bibfield  {author} {\bibinfo {author} {\bibfnamefont {C.~D.}\ \bibnamefont
  {Hill}}, \bibinfo {author} {\bibfnamefont {E.}~\bibnamefont {Peretz}},
  \bibinfo {author} {\bibfnamefont {S.~J.}\ \bibnamefont {Hile}}, \bibinfo
  {author} {\bibfnamefont {M.~G.}\ \bibnamefont {House}}, \bibinfo {author}
  {\bibfnamefont {M.}~\bibnamefont {Fuechsle}}, \bibinfo {author}
  {\bibfnamefont {S.}~\bibnamefont {Rogge}}, \bibinfo {author} {\bibfnamefont
  {M.~Y.}\ \bibnamefont {Simmons}}, \ and\ \bibinfo {author} {\bibfnamefont
  {L.~C.~L.}\ \bibnamefont {Hollenberg}},\ }\href {\doibase
  10.1126/sciadv.1500707} {\bibfield  {journal} {\bibinfo  {journal} {Science
  Advances}\ }\textbf {\bibinfo {volume} {1}} (\bibinfo {year} {2015}),\
  10.1126/sciadv.1500707}\BibitemShut {NoStop}%
\bibitem [{\citenamefont {Kalra}\ \emph {et~al.}(2014)\citenamefont {Kalra},
  \citenamefont {Laucht}, \citenamefont {Hill},\ and\ \citenamefont
  {Morello}}]{Kalra}%
  \BibitemOpen
  \bibfield  {author} {\bibinfo {author} {\bibfnamefont {R.}~\bibnamefont
  {Kalra}}, \bibinfo {author} {\bibfnamefont {A.}~\bibnamefont {Laucht}},
  \bibinfo {author} {\bibfnamefont {C.~D.}\ \bibnamefont {Hill}}, \ and\
  \bibinfo {author} {\bibfnamefont {A.}~\bibnamefont {Morello}},\ }\href@noop
  {} {\bibfield  {journal} {\bibinfo  {journal} {Phys. Rev. X}\ }\textbf
  {\bibinfo {volume} {4}},\ \bibinfo {pages} {021044} (\bibinfo {year}
  {2014})}\BibitemShut {NoStop}%
\bibitem [{\citenamefont {Batey}\ \emph {et~al.}(2014)\citenamefont {Batey},
  \citenamefont {Matthews},\ and\ \citenamefont {Patton}}]{Batey}%
  \BibitemOpen
  \bibfield  {author} {\bibinfo {author} {\bibfnamefont {G.}~\bibnamefont
  {Batey}}, \bibinfo {author} {\bibfnamefont {A.~J.}\ \bibnamefont {Matthews}},
  \ and\ \bibinfo {author} {\bibfnamefont {M.}~\bibnamefont {Patton}},\ }\href
  {http://stacks.iop.org/1742-6596/568/i=3/a=032014} {\bibfield  {journal}
  {\bibinfo  {journal} {Journal of Physics: Conference Series}\ }\textbf
  {\bibinfo {volume} {568}},\ \bibinfo {pages} {032014} (\bibinfo {year}
  {2014})}\BibitemShut {NoStop}%
\bibitem [{\citenamefont {Saraiva}\ \emph {et~al.}(2015)\citenamefont
  {Saraiva}, \citenamefont {Baena}, \citenamefont {Calderón},\ and\
  \citenamefont {Koiller}}]{Saraiva}%
  \BibitemOpen
  \bibfield  {author} {\bibinfo {author} {\bibfnamefont {A.~L.}\ \bibnamefont
  {Saraiva}}, \bibinfo {author} {\bibfnamefont {A.}~\bibnamefont {Baena}},
  \bibinfo {author} {\bibfnamefont {M.~J.}\ \bibnamefont {Calderón}}, \ and\
  \bibinfo {author} {\bibfnamefont {B.}~\bibnamefont {Koiller}},\ }\href
  {http://stacks.iop.org/0953-8984/27/i=15/a=154208} {\bibfield  {journal}
  {\bibinfo  {journal} {Journal of Physics: Condensed Matter}\ }\textbf
  {\bibinfo {volume} {27}},\ \bibinfo {pages} {154208} (\bibinfo {year}
  {2015})}\BibitemShut {NoStop}%
\bibitem [{\citenamefont {Witzel}\ \emph {et~al.}(2010)\citenamefont {Witzel},
  \citenamefont {Carroll}, \citenamefont {Morello}, \citenamefont
  {Cywi\ifmmode~\acute{n}\else \'{n}\fi{}ski},\ and\ \citenamefont
  {Das~Sarma}}]{Witzel}%
  \BibitemOpen
  \bibfield  {author} {\bibinfo {author} {\bibfnamefont {W.~M.}\ \bibnamefont
  {Witzel}}, \bibinfo {author} {\bibfnamefont {M.~S.}\ \bibnamefont {Carroll}},
  \bibinfo {author} {\bibfnamefont {A.}~\bibnamefont {Morello}}, \bibinfo
  {author} {\bibfnamefont {L.}~\bibnamefont {Cywi\ifmmode~\acute{n}\else
  \'{n}\fi{}ski}}, \ and\ \bibinfo {author} {\bibfnamefont {S.}~\bibnamefont
  {Das~Sarma}},\ }\href {\doibase 10.1103/PhysRevLett.105.187602} {\bibfield
  {journal} {\bibinfo  {journal} {Phys. Rev. Lett.}\ }\textbf {\bibinfo
  {volume} {105}},\ \bibinfo {pages} {187602} (\bibinfo {year}
  {2010})}\BibitemShut {NoStop}%
\bibitem [{\citenamefont {Kohn}\ and\ \citenamefont
  {Luttinger}(1955)}]{KohnLuttinger}%
  \BibitemOpen
  \bibfield  {author} {\bibinfo {author} {\bibfnamefont {W.}~\bibnamefont
  {Kohn}}\ and\ \bibinfo {author} {\bibfnamefont {J.~M.}\ \bibnamefont
  {Luttinger}},\ }\href {\doibase 10.1103/PhysRev.98.915} {\bibfield  {journal}
  {\bibinfo  {journal} {Phys. Rev.}\ }\textbf {\bibinfo {volume} {98}},\
  \bibinfo {pages} {915} (\bibinfo {year} {1955})}\BibitemShut {NoStop}%
\bibitem [{\citenamefont {Koiller}\ \emph {et~al.}(2004)\citenamefont
  {Koiller}, \citenamefont {Capaz}, \citenamefont {Hu},\ and\ \citenamefont
  {Das~Sarma}}]{KoillerDonorElectronWF}%
  \BibitemOpen
  \bibfield  {author} {\bibinfo {author} {\bibfnamefont {B.}~\bibnamefont
  {Koiller}}, \bibinfo {author} {\bibfnamefont {R.~B.}\ \bibnamefont {Capaz}},
  \bibinfo {author} {\bibfnamefont {X.}~\bibnamefont {Hu}}, \ and\ \bibinfo
  {author} {\bibfnamefont {S.}~\bibnamefont {Das~Sarma}},\ }\href {\doibase
  10.1103/PhysRevB.70.115207} {\bibfield  {journal} {\bibinfo  {journal} {Phys.
  Rev. B}\ }\textbf {\bibinfo {volume} {70}},\ \bibinfo {pages} {115207}
  (\bibinfo {year} {2004})}\BibitemShut {NoStop}%
\bibitem [{\citenamefont {Wellard}\ and\ \citenamefont
  {Hollenberg}(2005)}]{WellardHollenberg}%
  \BibitemOpen
  \bibfield  {author} {\bibinfo {author} {\bibfnamefont {C.~J.}\ \bibnamefont
  {Wellard}}\ and\ \bibinfo {author} {\bibfnamefont {L.~C.~L.}\ \bibnamefont
  {Hollenberg}},\ }\href {\doibase 10.1103/PhysRevB.72.085202} {\bibfield
  {journal} {\bibinfo  {journal} {Phys. Rev. B}\ }\textbf {\bibinfo {volume}
  {72}},\ \bibinfo {pages} {085202} (\bibinfo {year} {2005})}\BibitemShut
  {NoStop}%
\bibitem [{\citenamefont {Feher}(1959)}]{FeherENDOR}%
  \BibitemOpen
  \bibfield  {author} {\bibinfo {author} {\bibfnamefont {G.}~\bibnamefont
  {Feher}},\ }\href {\doibase 10.1103/PhysRev.114.1219} {\bibfield  {journal}
  {\bibinfo  {journal} {Phys. Rev.}\ }\textbf {\bibinfo {volume} {114}},\
  \bibinfo {pages} {1219} (\bibinfo {year} {1959})}\BibitemShut {NoStop}%
\bibitem [{\citenamefont {Hale}\ and\ \citenamefont {Mieher}(1969)}]{Hale}%
  \BibitemOpen
  \bibfield  {author} {\bibinfo {author} {\bibfnamefont {E.~B.}\ \bibnamefont
  {Hale}}\ and\ \bibinfo {author} {\bibfnamefont {R.~L.}\ \bibnamefont
  {Mieher}},\ }\href {\doibase 10.1103/PhysRev.184.739} {\bibfield  {journal}
  {\bibinfo  {journal} {Phys. Rev.}\ }\textbf {\bibinfo {volume} {184}},\
  \bibinfo {pages} {739} (\bibinfo {year} {1969})}\BibitemShut {NoStop}%
\bibitem [{\citenamefont {Klimov}\ \emph {et~al.}(2002)\citenamefont {Klimov},
  \citenamefont {S{\`a}nchez-Soto}, \citenamefont {Navarro},\ and\
  \citenamefont {Yustas}}]{Klimov}%
  \BibitemOpen
  \bibfield  {author} {\bibinfo {author} {\bibfnamefont {A.~B.}\ \bibnamefont
  {Klimov}}, \bibinfo {author} {\bibfnamefont {L.~L.}\ \bibnamefont
  {S{\`a}nchez-Soto}}, \bibinfo {author} {\bibfnamefont {A.}~\bibnamefont
  {Navarro}}, \ and\ \bibinfo {author} {\bibfnamefont {E.~C.}\ \bibnamefont
  {Yustas}},\ }\href {\doibase 10.1080/09500340210134675} {\bibfield  {journal}
  {\bibinfo  {journal} {Journal of Modern Optics}\ }\textbf {\bibinfo {volume}
  {49}},\ \bibinfo {pages} {2211} (\bibinfo {year} {2002})}\BibitemShut
  {NoStop}%
\bibitem [{\citenamefont {Laucht}\ \emph {et~al.}(2016)\citenamefont {Laucht},
  \citenamefont {Simmons}, \citenamefont {Kalra}, \citenamefont {Tosi},
  \citenamefont {Dehollain}, \citenamefont {Muhonen}, \citenamefont {Freer},
  \citenamefont {Hudson}, \citenamefont {Itoh}, \citenamefont {Jamieson},
  \citenamefont {McCallum}, \citenamefont {Dzurak},\ and\ \citenamefont
  {Morello}}]{Laucht}%
  \BibitemOpen
  \bibfield  {author} {\bibinfo {author} {\bibfnamefont {A.}~\bibnamefont
  {Laucht}}, \bibinfo {author} {\bibfnamefont {S.}~\bibnamefont {Simmons}},
  \bibinfo {author} {\bibfnamefont {R.}~\bibnamefont {Kalra}}, \bibinfo
  {author} {\bibfnamefont {G.}~\bibnamefont {Tosi}}, \bibinfo {author}
  {\bibfnamefont {J.~P.}\ \bibnamefont {Dehollain}}, \bibinfo {author}
  {\bibfnamefont {J.~T.}\ \bibnamefont {Muhonen}}, \bibinfo {author}
  {\bibfnamefont {S.}~\bibnamefont {Freer}}, \bibinfo {author} {\bibfnamefont
  {F.~E.}\ \bibnamefont {Hudson}}, \bibinfo {author} {\bibfnamefont {K.~M.}\
  \bibnamefont {Itoh}}, \bibinfo {author} {\bibfnamefont {D.~N.}\ \bibnamefont
  {Jamieson}}, \bibinfo {author} {\bibfnamefont {J.~C.}\ \bibnamefont
  {McCallum}}, \bibinfo {author} {\bibfnamefont {A.~S.}\ \bibnamefont
  {Dzurak}}, \ and\ \bibinfo {author} {\bibfnamefont {A.}~\bibnamefont
  {Morello}},\ }\href {\doibase 10.1103/PhysRevB.94.161302} {\bibfield
  {journal} {\bibinfo  {journal} {Phys. Rev. B}\ }\textbf {\bibinfo {volume}
  {94}},\ \bibinfo {pages} {161302} (\bibinfo {year} {2016})}\BibitemShut
  {NoStop}%
\end{thebibliography}
\end{document}